\documentclass[aps,pr,twocolumn,showpacs,superscriptaddress,groupedaddress,nofootinbib]{revtex4}  

\usepackage{longtable}
\usepackage{graphicx}  
\usepackage{ulem}   
\usepackage{comment} 
\usepackage{datetime}
\usepackage{dcolumn}

\usepackage{amsxtra}
\usepackage{euscript} 
\usepackage{bm}        
\usepackage{tabularx}
\usepackage{float}
\restylefloat{table}

\usepackage[cmtip,arrow]{xy}
\usepackage{pb-diagram,pb-xy}

\usepackage{amsmath}
\usepackage{amssymb}
\usepackage{amsthm}
\usepackage[pdftex]{color}
\usepackage[pdftex,colorlinks,citecolor=blue,linkcolor=blue,urlcolor=blue]{hyperref} 
\usepackage{graphicx}
\usepackage{dcolumn} 
\usepackage{bm} 

\usepackage{longtable}

\usepackage{ulem}   
\usepackage{comment} 
\usepackage{datetime}

\usepackage{tabularx}
\usepackage{float}
\restylefloat{table}

\newcommand{\beq}{\begin{equation}}
\newcommand{\eeq}{\end{equation}}
\newcommand{\bea}{\begin{eqnarray}}
\newcommand{\eea}{\end{eqnarray}}
\newcommand{\barr}{\begin{array}}
\newcommand{\earr}{\end{array}}

\long\def\begincomment#1\endcomment{}

\newcommand{\SU}{\mathrm{SU}}
\newcommand{\SO}{\mathrm{SO}}
\newcommand{\U}{\mathrm{U}}
\newtheorem{theorem}{Theorem}
\newtheorem{definition}{Definition}

\newtheorem{proposition}{Proposition}

\newcommand{\perm}{\mathrm{perm}}

\usepackage{pgf,tikz}
\usetikzlibrary{arrows}

\usepackage{bm}
\usepackage{bbold}

\usepackage{mathtools}

\DeclarePairedDelimiterX\braket[2]{\langle}{\rangle}{#1 \delimsize\vert #2}

\begin{document}


\title{   Ward-constrained melonic renormalization group flow
}

\author{Vincent Lahoche} \email{vincent.lahoche@cea.fr}   
\affiliation{Commissariat à l'\'Energie Atomique (CEA, LIST),
 8 Avenue de la Vauve, 91120 Palaiseau, France}

\author{Dine Ousmane Samary}
\email{dine.ousmanesamary@cipma.uac.bj}
\affiliation{International Chair in Mathematical Physics and Applications (ICMPA-UNESCO Chair), University of Abomey-Calavi,
072B.P.50, Cotonou, Republic of Benin}
\affiliation{Commissariat à l'\'Energie Atomique (CEA, LIST),
 8 Avenue de la Vauve, 91120 Palaiseau, France}

\date{\today}

\begin{abstract}
In recent years, some interesting investigations of the non-perturbative renormalization group equations for tensorial group field theories have been done in the truncation method, and completely discarding the Ward identities from their analysis. In this letter, in continuation of our recent series of papers, we present a new framework of investigation, namely the effective vertex expansion, allowing to consider infinite sectors rather than finite dimensional subspaces of the full theory space. We focus on the ultraviolet behavior, and provide a new and complete description of the renormalization group flow constrained with Ward identities. 
\end{abstract}


\maketitle

\section{Introduction} \label{sec1}
Functional renormalization group (FRG) applied to tensor models (TM) and group field theory (GFT) has been the subject of intense works in recent years because of its close relation with fluctuation problem of quantum gravity phenomenon \cite{Lahoche:2018hou}-\cite{Eichhorn:2018phj}.
Despite the  difficulties related to the nonlocal behavior of the  interactions  and combinatorics, some class of new technical computations are given to thank about the FRG  to tensorial  group field theory (TGFT) \cite{Lahoche:2018hou}-\cite{Lahoche:2018vun},\cite{Carrozza:2016tih}. First insights have been gained by nonperturbative Wetterich equation and, in particular by an investigation of the leading order melonic interactions, with a new method called effective vertex expansion (EVE) \cite{Lahoche:2018hou}-\cite{Lahoche:2018vun}. EVE  described the FRG without truncation as approximation and will certainly become a promising way in investigating nonperturbative field theory.  
 A lot of possible  phase transitions  which are identified  near the fixed point are shown to be non-physical due to the violation of the Ward identities (WI) \cite{Lahoche:2018ggd}. The WI is an additional constraint on the flow and  therefore should not be overlooked in the study of renormalization group.  In the symmetric phase we showed that, apart from the fact that no physical fixed point may be observed, the possible existence of first order phase transition in the reduced subspace of theory space can be given (see \cite{Lahoche:2018hou}).

A very useful concept for the study of the FRG and in particular the phase transitions is the coarse grained free energy or effective average action $\Gamma_k$. The $k$ dependence of this quantity is due to the regulator $r_k$ where $k$ ranges from IR to UV.  The nontrivial form of the Ward identity for the TGFT with nontrivial propagator in the functional actions is not a consequence of the regulator $r_k$ but rather is due to the violation of the kinetic term under $\U(N)$ symmetry. Let us remark that for standard gauge invariant theories like QED see \cite{Gies:2006wv} the regulator breaks generally the explicit invariance of the kinetic term and leads to a new non trivial Ward identity that depends on the regulator $r_k$. This is not the case for TGFT models for which the kinetic term intrinsically violates the $\U(N)$ symmetry. Therefore the appearance of  the regulator generalizes the definition of the theory but does not add any new information concerning the Ward identity.  Finally, the WI appears, like the flow equations themselves, as a formal consequence of the quantum model, and has to be taking into account on the same footing as the last ones.  As the Wetterich equation describes the $k$ variation of $\Gamma_k$ the Ward identity describes the momentum dependence of the same quantity.  Not to take into account this new dynamics related to the Ward identity  would be a serious  lack in the study of the FRG.

In this present letter, the  FRG is studied  with new alternative way by considering together the two dynamics of the average effective action. The first dynamic is given by the Wetterich flow equation  \cite{Wetterich:1991be}-\cite{Wetterich:1992yh} and the second by the Ward identity \cite{Lahoche:2018hou}.  We derived the melonic constraint flow by merging these two dynamics equations in the physical subspace $\mathcal{E}_{\mathcal C}$ of theory space. Note that the study of phase transitions is deeply related to the classification of all possible universality classes of the exact coincidence of the critical exponents. This universality is broken by the Ward constraint driving by the EVE like by the truncation method and such that the method proposed in this article  may be used for the  generalization to any other interaction of higher rank.

The paper is organized as follows.  In section \eqref{sec2} we provide in details useful ingredients for the description of the FRG to TGFT. In the section \eqref{EVE} the EVE is derived to thank about the FRG with a new alternative way without truncation. The corresponding flow equations which improves the truncation method are given. Section \eqref{melons} describes our new proposal to merge the Wetterich equation and the Ward identity in the physical melonic phase space $\mathcal{E}_{\mathcal C}$ of theory space. In the last section \eqref{conclusion} we give our conclusion.

\section{Preliminaries}\label{sec2}
A \textit{group field} $\varphi$ is a field, complex or real, defined over $d$--copies of a group manifold $\mathrm{G}$ rather than on space time:
\begin{equation}
\varphi: \mathrm{G}^d\to\mathbb{R},\mathbb{C}\,.
\end{equation}
Standard choices to make contact with physics are $\SU(2)$ and $\SO(4)$  \cite{Carrozza:2014rba}-\cite{Perez:2002vg}. In this paper, we focus only on the non-local aspects of the interactions, and consider the Abelian version of the theory, setting $\mathrm{G}=\U(1)$. For this choice, the field may be equivalently described on the Fourier dual group $\mathbb{Z}^d$ by a \textit{tensor field} $T:\mathbb{Z}^d\to \mathbb{C}$. We consider a theory for two complexes fields $\varphi$ and $\bar\varphi$, requiring two complex tensors fields $T$ and $\bar{T}$. The allowed configurations are then constrained by the choice of a specific action, completing the definition of the GFT. At the classical level, for free fields we choose the familiar form:
\begin{equation}
S_{\text{kin}}[T,\bar{T}]:=\sum_{\vec{p}\in\mathbb{Z}^d} \bar{T}_{p_1\cdots p_d}\left(\vec{p}\,^2+m^2\right) {T}_{p_1\cdots p_d}\,,\label{kin}
\end{equation}
with the standard notation $\vec{p}\,^2:=\sum_i p_i^2$, $\vec{p}:=(p_1,\cdots,p_d)$. For the rest of this paper we use the short notation $T_{\vec{p}}\equiv T_{p_1\cdots p_d}$. The equation \eqref{kin} defines the bare propagator $C^{-1}(\vec{p}\,):=\vec{p}\,^2+m^2$. Among the natural transformations that we can consider for a pair of complex tensor fields, the unitary transformations play an important role. They provide the principle that allows to build the interactions, which are chosen to be invariant under such a transformation. Denoting by $N$ the \textit{size} of the tensor field, restricting the domain of the indices $p_i$ into the window $[\![-N,N ]\!]$, we require invariance with respect to independent transformations along each of the $d$ indices of the tensors:
\begin{equation}
T^\prime_{p_1\cdots p_d}=\sum_{\vec{q}\in\mathbb{Z}^d} \left[\prod_{i=1}^d U^{(i)}_{p_iq_i}\right] \,T_{q_1\cdots q_d}\,,\label{unit}
\end{equation}
with $U^{(i)}(U^{(i)})^\dagger=\mathrm{id}$. Define $\mathbb{U}(N)$ as the set of unitary symmetries of size $N$, a transformation for tensors is then a set of $d$ independent elements of $\mathbb{U}(N)$, $\mathcal{U}:=(U_1,\cdots, U_d)\in\mathbb{U}(N)^d$, one per index of the tensor fields. The unitary symmetries admitting an inductive limit for arbitrary large $N$, we will implicitly consider the limit $N\to \infty$ in the rest of this paper. 
\noindent
We call \textit{bubble} all the invariant interactions which cannot be factorized into two or more smaller bubbles. Observe that because the transformations are independent, the bubbles are not local in the usual sense over the group manifold $\mathrm{G}^d$. However, locality does not make sense without physical content. In standard field theory for instance, or in physics in general, the locality is defined by the way following which the fields or particles interact together, and as for tensors, this choice reflects invariance with respect to some transformations like translations and rotations. With this respect, the transformation rule \eqref{unit} define both the nature of the field (a tensor) and the corresponding locality principle. To summarize:
\begin{definition}
Any interaction bubble is said to be local. By extension, any functions expanding as a sum of bubble will be said local. 
\end{definition}
This locality principle called \textit{traciality} in the literature has some good properties of the usual ones. In particular it allows to define local counter-terms and to follow the standard renormalization procedure for interacting quantum fields with UV divergences. In this paper, we focus on the quartic melonic model in rank $d=5$, describing by the classical interaction:
\begin{equation}
S_{\text{int}}[T,\bar T]= g \sum_{i=1}^d \vcenter{\hbox{\includegraphics[scale=0.8]{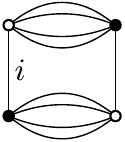} }}\,,\label{int}
\end{equation}
$g$ denoting the coupling constant and where we adopted the standard graphical convention \cite{Gurau:2011xp} to picture the interaction bubble as $d$-colored bipartite regular connected graphs. The black (resp. white) nodes corresponding to $T$ (resp. $\bar{T}$) fields, and the colored edges fixing the contractions of their indices. Note that, because we contract indices of the same color between $T$ and $\bar{T}$ fields, the unitary symmetry is ensured by construction. The model that we consider has been showed to be \textit{just renormalizable} is the usual sense, that is to say, all the UV  divergences can be subtracted with a finite set of counter-terms, for mass, coupling and field strength. From now on, we will consider $m^2$ and $g$ as the bare couplings, sharing their counter-terms, and we introduce explicitly the wave function renormalization $Z$ replacing the propagator $C^{-1}$ by 
\begin{equation}
C^{-1}(\vec{p}\,)=Z\vec{p}\,^2+m^2\,.
\end{equation}

\noindent
The equations \eqref{kin} and \eqref{int} define the classical model, without fluctuations. We quantize using path integral formulation, and define the partition function integrating over all configurations, weighted by $e^{-S}$:
\begin{equation}
\mathcal{Z}(J,\bar J):=\int dT d\bar T \,e^{-S[T,\bar T]+\langle \bar{J},T\rangle+\langle \bar{T},J\rangle}\,,\label{quantum}
\end{equation}
the sources being tensor fields themselves $J,\bar J:\mathbb{Z}^d\to \mathbb{C}$ and $\langle \bar{J},T\rangle:=\sum_{\vec{p}}\bar{J}_{\vec{p}}\, T_{\vec{p}}$. Note that the quantization procedure provide a canonical definition of what is UV and what is IR. The UV theory corresponding to the classical action $S=S_{\text{kin}}+S_{\text{int}}$ whereas the IR theory corresponds to the standard effective action defined as the Legendre transform of the free energy $\mathcal{W}:=\ln(\mathcal{Z}(J,\bar J))$. \\

\noindent
Renormalization in standard field theory allows to subtract divergences, and it has been showed that quantum GFT can be renormalized in the usual sense \cite{BenGeloun:2011rc}-\cite{BenGeloun:2012pu}. With respect to the quantization procedure moreover, the renormalization group allows to describe quantum effects \textit{scale by scale}, through more and more effective models, defining a path from UV to IR by integrating out fluctuation of increasing size. \\

\noindent
Recognizing this path from UV to IR as an element of the quantization procedure itself, we substitute to the global quantum description \eqref{quantum} a set of models $\{\mathcal{Z}_k\}$ indexed by a referent scale $k$. This scale define what is UV, and integrated out and what is IR, and frozen out from the long distance physics. The set of scales may be discrete or continuous, and in this paper we choose a continuous description $k\in [0,\Lambda]$ for some fundamental UV cut-off $\Lambda$. There are several ways to build what we call functional renormalization group . We focus on the Wetterich-Morris approach  \cite{Wetterich:1991be}-\cite{Wetterich:1992yh}, $ \mathcal{Z}_k(J,\bar J)$ being defined as:
\begin{equation}
\mathcal{Z}_k(J,\bar J):=\int dT d\bar T \,e^{-S_k[T,\bar T]+\langle \bar{J},T\rangle+\langle \bar{T},J\rangle}\,,\label{quantum2}
\end{equation}
with: $S_k[T,\bar T]:=S[T,\bar T]+\sum_{\vec{p}}\,\bar{T}_{\vec{p}}\, r_k(\vec{p}\,^2)T_{\vec{p}}$. The momentum dependent mass term $r_k(\vec{p}\,^2)$ called \textit{regulator} vanish for UV fluctuations $\vec{p}\,^2\gg k^2$ and becomes very large for the IR ones $\vec{p}\,^2\ll k^2$. Some additional properties for $r_k(\vec{p}\,^2)$ may be found in standard references \cite{Litim:2000ci}-\cite{Litim:2001dt}. Without explicit mentions, we focus on the Litim's modified regulator:
\begin{equation}
r_k(\vec{p}\,^2):=Z(k)(k^2-\vec{p}\,^2)\theta(k^2-\vec{p}\,^2)\,,\label{regulator}
\end{equation}
where $\theta$ designates the Heaviside step function and $Z(k)$ is the running wave function strength. The renormalization group flow equation, describing the trajectory of the RG flow into the full theory space is the so called Wetterich equation  \cite{Wetterich:1991be}-\cite{Wetterich:1992yh}, which for our model writes as:
\begin{equation}
\frac{\partial}{\partial k} \Gamma_k= \sum_{\vec{p}} \frac{\partial r_k}{\partial k}(\vec{p}\,) \left( \Gamma_k^{(2)}+r_k \right)^{-1}_{\vec{p}\,\vec{p}}\,,\label{Wett}
\end{equation}
where $(\Gamma_k^{(2)})_{\vec{p}\,\vec{p}\,^\prime}$ is the second derivative of the \textit{average effective action} $\Gamma_k$ with respect to the classical fields $M$ and $\bar{M}$:
\begin{equation}
\left(\Gamma_k^{(2)}\right)_{\vec{p}\,\vec{p}\,^\prime}=\frac{\partial^2\Gamma_k}{\partial M_{\vec{p}}\,\partial \bar{M}_{\vec{p}\,^\prime}}\,,
\end{equation}
where $M_{\vec{p}}=\partial \mathcal{W}_k/\partial\bar{J}_{\vec{p}}$, $\bar M_{\vec{p}}=\partial \mathcal{W}_k/\partial{J}_{\vec{p}}$ and:
\begin{align}
\nonumber\Gamma_k[M,\bar M]+\sum_{\vec{p}}\,\bar{M}_{\vec{p}}\, r_k(\vec{p}\,^2)M_{\vec{p}}&:=\langle \bar{M},J\rangle+\langle \bar J,M\rangle\\
&\,-\mathcal{W}_k(M,\bar M)\,,
\end{align}
with $\mathcal{W}_k=\ln(\mathcal{Z}_k)$.\\

\noindent
The flow equation \eqref{Wett} is a consequence of the variation of the propagator, indeed
\begin{equation}
\frac{\partial r_k}{\partial k} =\frac{\partial C^{-1}_k}{\partial k} \,,
\end{equation}
for the \textit{effective covariance} $C^{-1}_k:=C^{-1}+r_k$. But the propagator has other source of variability. In particular, it is not invariant with respect to the unitary symmetry of the classical interactions \eqref{int}. Focusing on an infinitesimal transformation : $\delta_1:=(\mathrm{id}+\epsilon,\mathrm{id},\cdots,\mathrm{id})$ acting non-trivially only on the color $1$ for some infinitesimal anti-hermitian transformations $\epsilon$, the transformation rule for the propagator follows the Lie bracket:
\begin{equation}
\mathcal{L}_{\delta_1} C^{-1}_k= [C^{-1}_k,\epsilon]\,. 
\end{equation}
The source terms are non invariant as well. However, due to the translation invariance of the Lebesgue measure $dT d\bar T$ involved in the path integral \eqref{quantum2}, we must have $\mathcal{L}_{\delta_1} \mathcal{Z}_k=0$. Translating this invariance at the first order in $\epsilon$ provide a non-trivial \textit{Ward-Takahashi identity} for the quantum model:
\begin{theorem} \textbf{(Ward identity.)}
The non-ivariance of the kinetic action with respect to unitary symmetry induce non-trivial relations between $\Gamma^{(n)}$ and $\Gamma^{(n+2)}$ for all $n$, summarized as:
\begin{align}
\nonumber\sum_{\vec{p}_\bot, \vec{p}_\bot^{\prime}}{\vphantom{\sum}}'
&\bigg\{\big[C_k^{-1}(\vec{p})-C_k^{-1}(\vec{p}\,^{\prime})\big]\left[\frac{\partial^2 \mathcal{W}_k}{\partial \bar{J}_{\vec{p}\,^\prime}\,\partial {J}_{\vec{p}}}+\bar{M}_{\vec{p}}M_{\vec{p}\,^\prime}\right]\\
&\quad-\bar{J} _{\vec{p}}\,M_{\vec{p}\,^\prime}+{J} _{\vec{p}\,^\prime}\bar{M}_{\vec{p}}\bigg\}=0\,.\label{Ward0}
\end{align}
where $\sum_{\vec{p}_\bot, \vec{p}_\bot^{\prime}}' :=\sum_{\vec{p}_\bot, \vec{p}_\bot^{\prime}} \delta_{\vec{p}\,\vec{p}_\bot^\prime}$.
\end{theorem}
In this statement we introduced the notations $\vec{p}_\bot:=(p_2,\cdots,p_d)\in\mathbb{Z}^{d-1}$ and $\delta_{\vec{p}\,\vec{p}_\bot^\prime}=\prod_{j\neq 1}\, \delta_{p_j\,p_j^\prime}$.   Equations \eqref{Wett} and \eqref{Ward0} are two formal consequences of the path integral \eqref{quantum2}, coming both from the non-trivial variations of the propagator. Therefore, there are no reason to treat these two equations separately. This formal proximity is highlighted in their expanded forms, comparing equations \eqref{florence}--\eqref{florence2} and \eqref{W1}--\eqref{W2}. Instead of a set of partition function, the quantum model may be alternatively defined as an (infinite) set of effective vertices $\mathcal{Z}_k\sim \{\Gamma^{(n)}_k\}=:\mathfrak{h}_k$.  RG equations dictate how to move from $\mathfrak{h}_k\underset{\text{RG}}{ \to}\mathfrak{h}_{k+\delta k}$ whereas Ward identities dictate how to move in the momentum space, along $\mathfrak{h}_k$. 

\section{Effective vertex expansion}\label{EVE}
This section essentially summarize the state of the art in \cite{Lahoche:2018hou}-\cite{Lahoche:2018vun}. The exact RG equation cannot be solved except for very special cases. The main difficulty is that the Wetterich equation \eqref{Wett} split as a infinite hierarchical system, the derivative of $\Gamma^{(n)}$ involving $\Gamma^{(n+2)}$, and so one. Appropriate approximation schemes are then required to extract an information on the exact solutions. The effective vertex expansion (EVE) is a recent technique allowing to build an approximation considering infinite sectors rather than crude truncations on the full theory space. We focus on the \textit{melonic sector}, sharing all the divergences of the model and then dominating the flow in the UV. One recall that melonic diagrams are defined as the diagram with optimal degree of divergence. Fixing some fundamental cut-off $\Lambda$, we consider the domain $\Lambda \ll k \ll 1$, so far from the deep UV and the deep IR regime. At this time, the flow is dominated by the renormalized couplings, have positive or zero \textit{flow dimension} (see \cite{Lahoche:2018oeo}). We recall that the flow dimension reflect the behavior of the RG flow of the corresponding quantity, and discriminate between essential, marginal and inessential couplings just like standard dimension in quantum field theory\footnote{For ordinary quantum field theory, the dimension is fixed by the background itself. Without background, this is the behavior of the RG flow which fix the canonical dimension.}. Because our theory is just-renormalizable, one has necessarily $[m^2]=2$ and $[g]=0$, denoting as $[X]$ the flow dimension of $X$. \\

\noindent
The basic strategy of the EVE is to close the hierarchical system coming from \eqref{Wett} using the analytic properties of the effective vertex functions\footnote{Melonic diagrams may be easily counted as "trees", and the (renormalized) melonic perturbation series is easy to sum. } and the rigid structure of the melonic diagrams. More precisely, the EVE express all the melonic effective vertices $\Gamma^{(n)}$ having negative flow dimension (that is for $n>4$) in terms of effective vertices with positive or null flow dimension, that is $\Gamma^{(2)}$ and $\Gamma^{(4)}$, and their flow is entirely drove by just-renormalizable couplings. As recalled, in this way we keep the entirety of the melonic sector and the full momentum dependence of the effective vertices. \\

\noindent
We work into the \textit{symmetric phase}, i.e. in the interior of the domain where the vacuum $M=\bar{M}=0$ make sense. This condition ensure that effective vertices with an odd number of external points, or not the same number of black and white external nodes have to be discarded from the analysis. These ones being called \textit{assorted functions}. Moreover, due to the momentum conservation along the boundaries of faces, $\Gamma^{(2)}_k$ must be diagonal: 
\begin{equation}
\Gamma^{(2)}_{k,\,\vec{p}\,\vec{q}}=\Gamma_k^{(2)}(\vec{p}\,)\delta_{\vec{p}\,\vec{q}}\,.
\end{equation}
We denote as $G_k$ the effective $2$-point function $G^{-1}_k:=\Gamma_k^{(2)}+r_k$. \\

\noindent
The main assumption of the EVE approach is the existence of a finite analyticity domain for the leading order effective vertex functions, in which they may be identified with the resumed perturbative series. For the melonic vertex function, the existence of a such analytic domain is ensured, melons can be mapped as trees and easily summed. Moreover, these resumed functions satisfy the Ward-Takahashi identities, written without additional assumption than cancellation of odd ans assorted effective vertices. One then expect that the symmetric phase entirely cover the perturbative domain. \\

\begin{figure}
\includegraphics[scale=0.4]{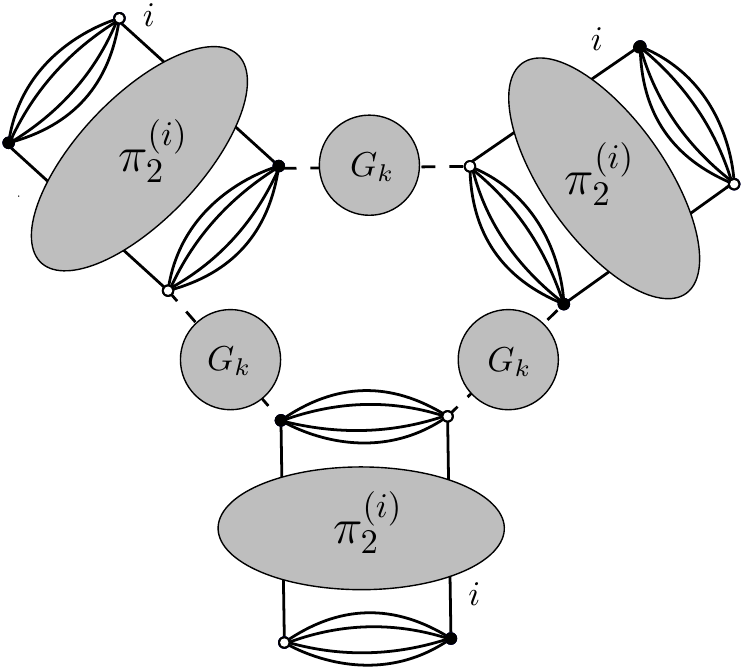} 
\caption{Interne structure of the 1PI $6$-points graphs. } \label{fig6point}
\end{figure}

\noindent
Among the properties of the melonic diagrams, we recall the following statement:
\begin{proposition}\label{propmelons}
Let $\mathcal{G}_N$  be a $2N$-point 1PI melonic diagrams build with more than one vertices for a purely quartic melonic model. We call external vertices the vertices hooked to at least one external edge of $\mathcal{G}_N$ has :
\begin{itemize}
\item   Two external edges per external vertices, sharing $d-1$ external faces of length one. 
\item  $N$ external faces of the same color running through the interior of the diagram. 
\end{itemize}
\end{proposition}
Due to this proposition, the melonic effective vertex $\Gamma^{(n)}_k$  decompose as $d$ functions $\Gamma^{(n),i}_k$, labeled with a color index $i$:
\begin{equation}
\Gamma^{(n)}_{k,\,\vec{p}_1,\cdots,\vec{p}_n}=\sum_{i=1}^d \Gamma^{(n),i}_{k,\,\vec{p}_1,\cdots,\vec{p}_n}\,.
\end{equation}
The Feynman diagrams contributing to the perturbative expansion of $\Gamma^{(n,i)}_{k,\,\vec{p}_1,\cdots,\vec{p}_n}$ fix the relations between the different indices. For $n=4$ for instance, we get, from proposition \ref{propmelons}:
\begin{equation}
\Gamma_{\vec{p}_1,\vec{p}_2,\vec{p}_3,\vec{p}_4}^{(4),i} = \vcenter{\hbox{\includegraphics[scale=0.5]{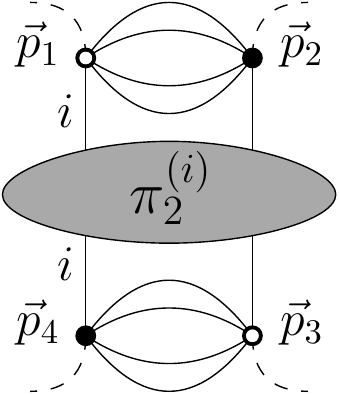} }}+\, \vcenter{\hbox{\includegraphics[scale=0.5]{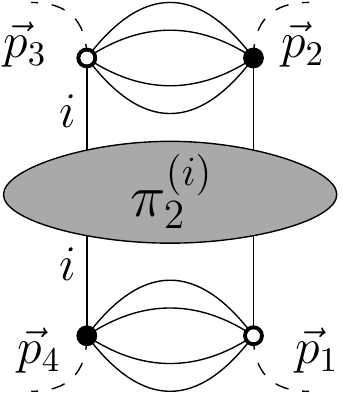} }}\,,\label{decomp4}
\end{equation}
Where  the half dotted edges correspond to the amputated external propagators, and the reduced vertex functions $\pi_2^{(i)}:\mathbb{Z}^2\to \mathbb{R}$ denotes the sum of the interiors of the graphs, excluding the external nodes and the colored edges hooked to them. In the same way, one expect that the melonic effective vertex $\Gamma_{\text{melo\,}\vec{p}_1,\vec{p}_2,\vec{p}_3,\vec{p}_4,\vec{p}_5,\vec{p}_6}^{(6),i}$ is completely determined by a reduced effective vertex $\pi_3^{(i)}:\mathbb{Z}^3\to\mathbb{R}$ hooked to a boundary configuration such as:
\begin{equation}
\Gamma_{\vec{p}_1,\vec{p}_2,\vec{p}_3,\vec{p}_4,\vec{p}_5,\vec{p}_6}^{(6),i}=\vcenter{\hbox{\includegraphics[scale=0.4]{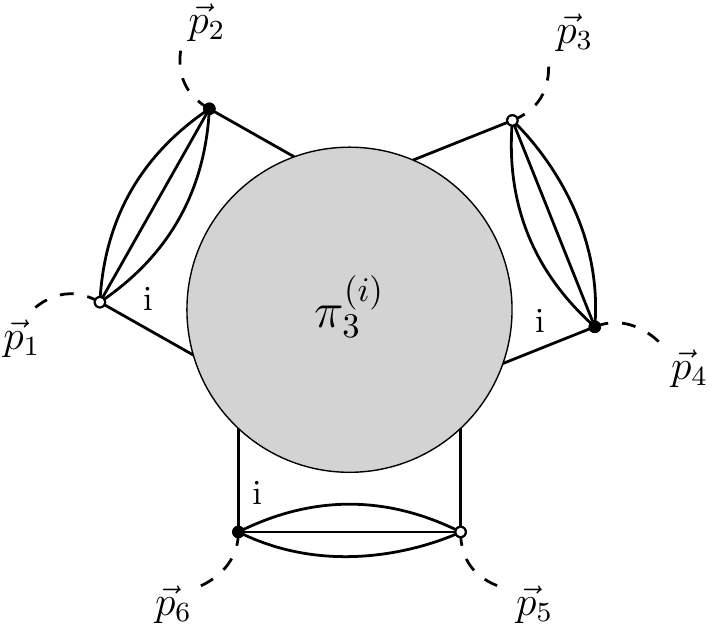} }}\,\,\, +\,\perm\,,
\end{equation}
and so one for $\Gamma^{(n),i}_{k,\,\vec{p}_1,\cdots,\vec{p}_n}$, involving the reduced vertex  $\pi_n^{(i)}:\mathbb{Z}^n\to\mathbb{R}$. In the last expression, $\perm$ denote the permutation of the external edges like in \eqref{decomp4}. The reduced vertices $\pi_2^{(i)}$ can be formally resumed as a geometric series \cite{Lahoche:2018hou}-\cite{Lahoche:2018oeo}:
\begin{align}
\nonumber \pi_{2,pp}^{(1)}&= 2\left(g-2g^2 \mathcal{A}_{2,p}+4g^3(\mathcal{A}_{2,p})^2-\cdots\right)\\
&=\frac{2g}{1+2g\mathcal{A}_{2,p}}\,,\label{eff4}
\end{align}
where $ \pi_{2,pp}^{(1)}$ is the diagonal element of the matrix $ \pi_{2}^{(1)}$ and :
\begin{equation}
\mathcal{A}_{n,p}:=\sum_{\vec{p}} \,G_k^n(\vec{p})\delta_{p\,p_1}\,.
\end{equation}
The reduced vertex $\pi_{2,pp}^{(1)}$ depend implicitly on $k$, and the renormalization conditions defining the \textit{renormalized coupling} $g_r$ are such that:
\begin{equation}
\pi_{2,00}^{(i)}\vert_{k=0}=:2g_r\,.
\end{equation}
For arbitrary $k$, the zero momentum value of the reduced vertex define the effective coupling for the local quartic melonic interaction: $\pi_{2,00}^{(i)}=:2g(k)$. The explicit expression for $\pi_3^{(1)}$ may be investigated from the proposition\ref{propmelons}. The constraint over the boundaries and the recursive definition of melonic diagram enforce the internal structure pictured on Figure \ref{fig6point} below [see Lahoche-Samary]. Explicitly:
\begin{equation}
\pi_{3,ppp}^{(i)}=(\pi_{2,pp}^{(i)})^3\,\mathcal{A}_{3,p}\,, \label{6pp}
\end{equation}
The two orientations of the external effective vertices being took into account in the definition of $\pi_{2,pp}^{(i)}$.
Expanding the exact flow equation \eqref{Wett}, and keeping only the relevant contraction for large $k$, one get the following relevant contributions for $\dot\Gamma_{k}^{(2)}$ and $\dot\Gamma_{k}^{(4)}$ :
\begin{equation}
\dot\Gamma_{k}^{(2)}= -\,\sum_{i=1}^d\,\vcenter{\hbox{\includegraphics[scale=0.65]{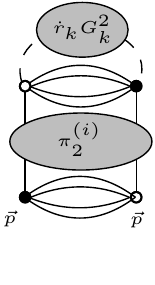} }}\label{florence}
\end{equation}
\begin{equation}
\dot\Gamma_{k}^{(4),i}=-\,2\,\vcenter{\hbox{\includegraphics[scale=0.45]{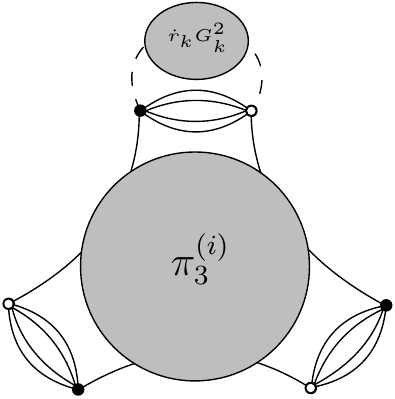} }}+\,\,8\,\vcenter{\hbox{\includegraphics[scale=0.45]{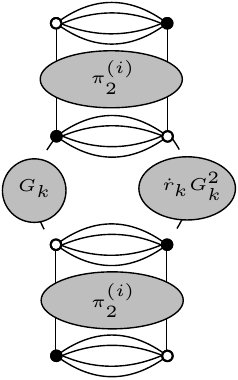} }}\label{florence2}
\end{equation}
where $\dot X:=k \partial X/\partial k$. The computation require the explicit expression of $\Gamma^{(2)}_k$. In the melonic sector, the self energy obey to a closed equation, reputed difficult to solve. We approximate the exact solution by considering only the first term in the derivative expansion in the interior of the windows of momenta allowed by $\dot r_k$:
\begin{equation}
\Gamma_{k}^{(2)}(\vec{p}\,):=Z(k)\vec{p}\,^2+m^2(k)\,,\label{derivexp}
\end{equation}
where $Z(k):=\partial\Gamma_{k}^{(2)}/\partial p_1^2(\vec{0}\,)$ and $m^2(k):=\Gamma_{k}^{(2)}(\vec{0}\,)$ are both renormalized and effective field strength and mass. From the definition \eqref{regulator}, and with some calculation (see \cite{Lahoche:2018oeo}), we obtain the following statement:
\begin{proposition}
In the UV domain $\Lambda\ll k \ll 1$ and in the symmetric phase, the leading order flow equations for essential and marginal local couplings are given by: 
\begin{align}
\left\{
    \begin{array}{ll}
       \beta_m&=-(2+\eta)\bar{m}^{2}-10 \bar{g}\,\frac{\pi^2}{(1+\bar{m}^{2})^2}\,\left(1+\frac{\eta}{6}\right)\,, \\
       \beta_{g}&=-2\eta \bar{g}+4\bar{g}^2 \,\frac{\pi^2}{(1+\bar{m}^{2})^3}\,\left(1+\frac{\eta}{6}\right)\Big[1\\
       &\quad-\frac{1}{2}\pi^2\bar{g}\left(\frac{1}{(1+\bar{m}^{2})^2}+\left(1+\frac{1}{1+\bar{m}^{2}}\right)\right)\Big]\,. \label{syst3}
    \end{array}
\right.
\end{align}
With:
\begin{equation}
\eta=4\bar{g}\pi^2\frac{(1+\bar{m}^{2})^2-\frac{1}{2}\bar{g}\pi^2(2+\bar{m}^{2})}{(1+\bar{m}^{2})^2\Omega(\bar{g},\bar{m}^{2})+\frac{(2+\bar{m}^{2})}{3}\bar{g}^2\pi^4}\,,\label{eta1}
\end{equation}
and
\begin{equation}
\Omega(\bar m^2,\bar g):=(\bar m^2+1)^2-\pi^2\bar g\,.
\end{equation}
\end{proposition}
Where in this proposition $\beta_g:=\dot{\bar g}$, $\beta_m:=\dot{\bar{m}}^2$ and the effective-renormalized mass and couplings are defined as: $\bar{g}:=Z^{-2}(k)g(k)$ and $\bar{m}^2:=Z^{-1}(k)k^{-2}m^2(k)$. For the computation, note that we made use of the approximation \eqref{derivexp} only for absolutely convergent quantities, and into the windows of momenta allowed by $\dot{r}_k$. As pointed out in \cite{Lahoche:2018hou}-\cite{Lahoche:2018oeo}, taking into account the full momentum dependence of the effective vertex $\pi_2^{(i)}$ in \eqref{eff4} drastically modify the expression of the anomalous dimension $\eta$ with respect to crude truncation. In particular, the singularity line discussed in \cite{Lahoche:2018ggd} disappears below the singularity $\bar{m}^2=-1$. Moreover, because all the effective melonic vertices only depend on $\bar{m}^2$ and $\bar{g}$, any fixed point for the system \eqref{syst3} is a global fixed point for the melonic sector. The system \eqref{syst3} admits a fixed point for $p:=(\bar{g}_*;\bar{m}^2_*)\approx(0.003;-0.55)$. 

\section{The melonic constrained flow}\label{melons}
To close the hierarchical system derived from \eqref{Wett} and obtain the autonomous set \eqref{syst3}, we made use of the explicit expressions \eqref{eff4} and \eqref{6pp}. In this derivation we mentioned the Ward identity but they do not contribute explicitly. In this section we take into account their contribution, and show that they introduce a strong constraint over the RG trajectories. \\

\noindent
Deriving successively the Ward identity \eqref{Ward0} with respect to external sources, and setting $J=\bar{J}=0$ at the end of the computation, we get the two following relations involving $\Gamma_k^{(4)}$ and $\Gamma_k^{(6)}$ (see \cite{Lahoche:2018ggd})
\begin{equation}
\pi_{2,00}^{(1)}\,\mathcal{L}_{2,k}=-\frac{\partial}{\partial p_1^2}\left(\Gamma_k^{(2)}(\vec{p}\,)-Z\vec{p}\,^2\right)\big\vert_{\vec{p}=\vec{0}}\,, \label{W1}
\end{equation}
\begin{equation}
2\left(\pi_{3,00}^{(1)}\,\mathcal{L}_{2,k}-(\pi_{2,00}^{(1)})^2\,\mathcal{L}_{3,k}\right)=-\frac{d}{dp_1^2}\pi_{2,p_1p_1}^{(1)}\big\vert_{p_1=0}\,,\label{W2}
\end{equation}
where:
\begin{equation}
\mathcal{L}_{n,k}:= \sum_{\vec{p}_\bot}\left(Z+\frac{\partial r_k}{\partial p_1^2} (\vec{p}_\bot)\right)G^n_k(\vec{p}_\bot)\,.
\end{equation}
It can be easily checked that the structure equations \eqref{eff4} and \eqref{6pp} satisfy the second Ward identity \eqref{W2} see \cite{Lahoche:2018hou}-\cite{Lahoche:2018oeo} and also  \cite{Samary:2014tja}-\cite{Sanchez:2017gxt}. In the same way the first Ward identity  \eqref{W1} has been checked to be compatible with the equation \eqref{eff4} and the melonic closed equation for the $2$-point function. However, the last condition 
does not exhaust the information contained in \eqref{eff4}. Indeed, with the same level of approximation as for the computation of the flow equations \eqref{syst3}, the first Ward identity can be translated locally as a constraint between beta functions (see \cite{Lahoche:2018oeo}):
\begin{equation}
\mathcal{C}:=\beta_g+\eta\bar{g}\, \frac{\Omega(\bar{g},\bar{m}^2)}{(1+\bar{m}^2)^2}-\frac{2\pi ^2\bar{g}^2}{(1+\bar{m}^2)^3}\beta_m=0\,.\label{const}
\end{equation}
Generally, the solutions of the system \eqref{syst3} do not satisfy the constraint $\mathcal{C}=0$. We call \textit{physical melonic phase space} and denote as $\mathcal{E}_{\mathcal{C}}$ the subspace of the melonic  theory space satisfying $\mathcal{C}=0$. A attempt to describe this space has been provided in \cite{Lahoche:2018hou}. In particular, we showed that there are no global fixed point of \eqref{syst3} which satisfy the constraint $\mathcal{C}=0$. \\

\noindent
In the description of the physical flow over $\mathcal{E}_{\mathcal{C}}$ provided in \cite{Lahoche:2018hou}, we substituted the explicit expressions of $\beta_g$, $\beta_m$ and $\eta$, translating the relations between velocities as a complicated constraint on the couplings $\bar{g}$ and $\bar{m}^2$. Solving this constraint, we build a systematic projection of the RG trajectories. Beyond the fact that this strategy is difficult to extend for renormalized models involving higher order interactions, even for the quartic molonic model some difficulty appear, as multi-branch phenomenon \cite{Lahoche:2018hou}. In this section we provide an alternative description which simplify the description of $\mathcal{E}_{\mathcal{C}}$ and can be easily extended for model with higher order interactions. Substituting the flow equations \eqref{syst3} into the constraint \eqref{const}, we implicitly impose the conservation of the relation \eqref{6pp} between $\pi_2^{(i)}$ and $\pi_3^{(i)}$ on $\mathcal{E}_{\mathcal{C}}$. We propose to relax this constraint, fixing $\pi_3^{(i)}$ by the flow itself. Our procedure is the following. \\

\noindent
(1) We keep $\beta_m$ and fix $\beta_g$ from the equation \eqref{const}:
\begin{equation}
\left\{
    \begin{array}{ll}
        \beta_m&=-(2+\eta)\bar{m}^{2}-\,\frac{10\pi^2\bar{g}}{(1+\bar{m}^{2})^2}\,\left(1+\frac{\eta}{6}\right)\,,\\
       \beta_g&=-\eta\bar{g}\, \frac{\Omega(\bar{g},\bar{m}^2)}{(1+\bar{m}^2)^2}+\frac{2\pi ^2\bar{g}^2}{(1+\bar{m}^2)^3}\beta_m\,.
    \end{array}
\right.\label{systprime}
\end{equation}
(2) We fix $\pi_{3,00}^{(i)}$ dynamically from the flow equation \eqref{florence2}:
\begin{align}
\nonumber\beta_g=-2\eta \bar{g} &-\frac{1}{2} \bar{\pi}_3^{(1)}\frac{\pi^2}{(1+\bar{m}^{2})^2}\left(1+\frac{\eta}{6}\right)\\
& +4\bar{g}^2 \,\frac{\pi^2}{(1+\bar{m}^{2})^3}\left(1+\frac{\eta}{6}\right)\label{pi3dyn}
\end{align}
(3) We compute $\frac{d}{dp_1^2}\pi_{2,00}^{(i)}$ from equation \eqref{W2}, and finally deduce the anomalous dimension $\eta$. The computation require the sums $\mathcal{L}_{2,k}$ and $\mathcal{L}_{3,k}$. Following \cite{Lahoche:2018hou}-\cite{Lahoche:2018oeo}, $\mathcal{L}_{3,k}$ may be computed using the approximation \eqref{derivexp}, but not $\mathcal{L}_{2,k}$ which has a vanishing power counting. However, $\mathcal{L}_{2,k}$ may be expressed in term of $Z(k)$ and $g(k)$ from equation \eqref{W1}. Indeed, setting $k=0$ and fixing the renormalization condition such that $Z(k=0)=1$\footnote{This condition may be refined, see [Lahoche-Samary 2], but this point has no consequence on our discussion.}, we get that, in the continuum limit $\Lambda\to \infty$, $Z\to 0$.  To summarize \eqref{W1} reduces to $-2g(k)\mathcal{L}_{2,k}=Z(k)$, and from \eqref{W2}:
\begin{equation}
\frac{d}{dp_1^2}\pi_{2,00}^{(1)}=\left(Z(k)\frac{\pi_{3,00}^{(1)}}{g(k)}+2(\pi_{2,00}^{(1)})^2\,\mathcal{L}_{3,k}\right)\,. \label{deriv2}
\end{equation}
Computing $\mathcal{L}_{3,k}$, one get straightforwardly, in the continuum limit:
\begin{equation}
\mathcal{L}_{3,k}=-\frac{1}{2Z^2(k)k^2}\frac{\pi^2}{(1+\bar{m}^2)^3}\,.
\end{equation}
Then, from equations \eqref{deriv2}, \eqref{pi3dyn} and from the flow equation \eqref{florence}, it is easy to get the explicit expression of $\eta$ on $\mathcal{E}_{\mathcal{C}}$, replacing the expression \eqref{eta1}:
\begin{equation}
\eta=4\pi^2\bar{g}\frac{(1+\bar{m}^2)^3+9\pi^2\bar{g}}{(1+\bar{m}^2)^5-\Omega^\prime(\bar{g},\bar{m}^2)}\,,\label{etaprime}
\end{equation}
with:
\begin{equation}
\Omega^\prime(\bar{g},\bar{m}^2) := \pi^2\bar{g}(1+\bar{m}^2)^3\left(1-\frac{8}{3(1+\bar{m}^2)^2}\right)+6\pi^4\bar{g}^2
\end{equation}

\noindent
Note that the two equations \eqref{W2} and \eqref{const} are satisfied by construction. Moreover, the hierarchy remains closed. Indeed, $\pi_{3,00}^{(i)}$ being fixed, we may compute $\dot{\pi}_{3,00}^{(i)}$ and to equal with the corresponding frow equation provided by \eqref{Wett}. We then fix $\pi_{4,00}^{(i)}$, and so one. \\

\begin{figure}
$\underset{a}{\vcenter{\hbox{\includegraphics[scale=0.5]{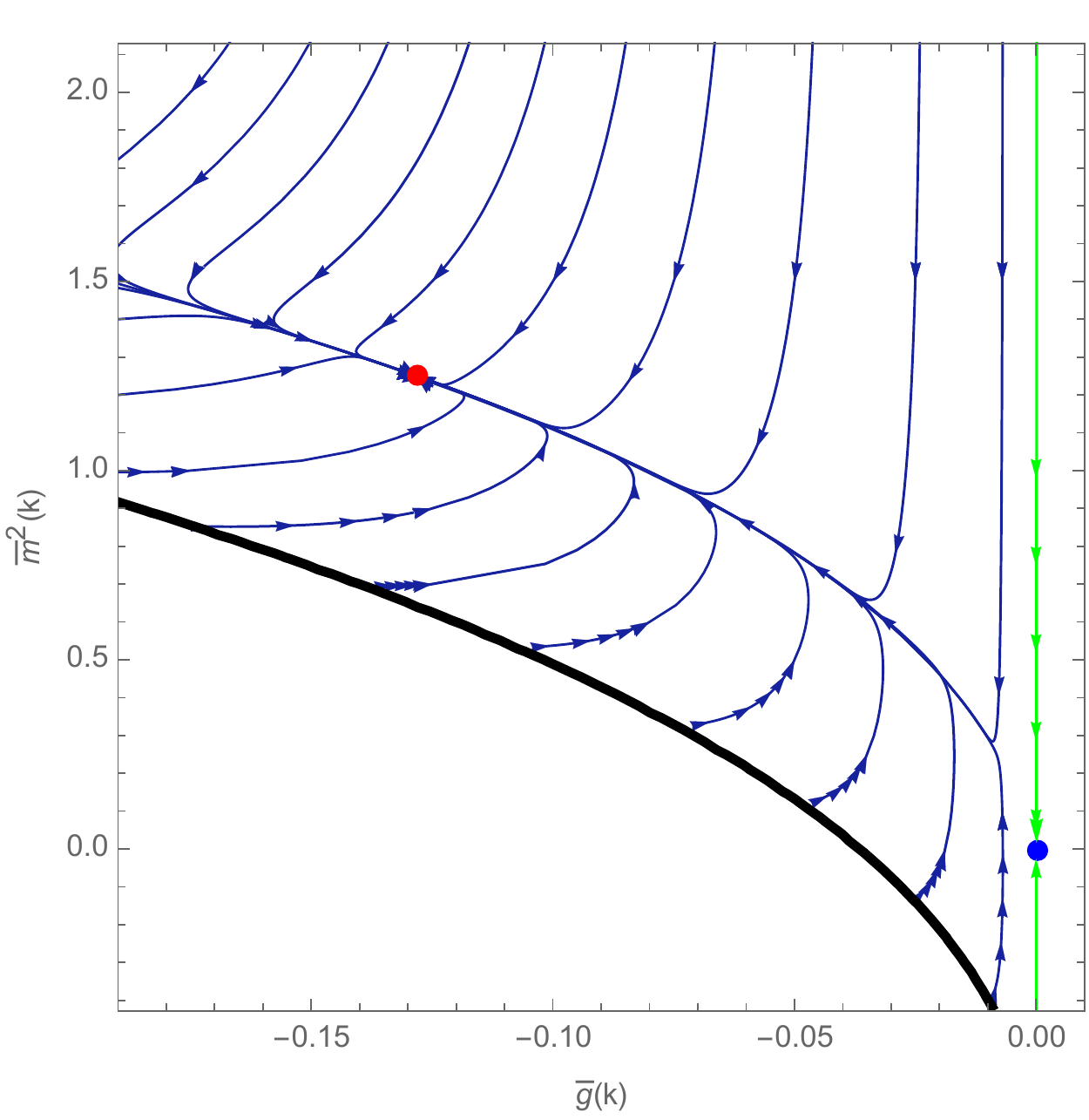} }}}$\\
$\vspace{0.5cm}$
$\underset{b}{\vcenter{\hbox{\includegraphics[scale=0.5]{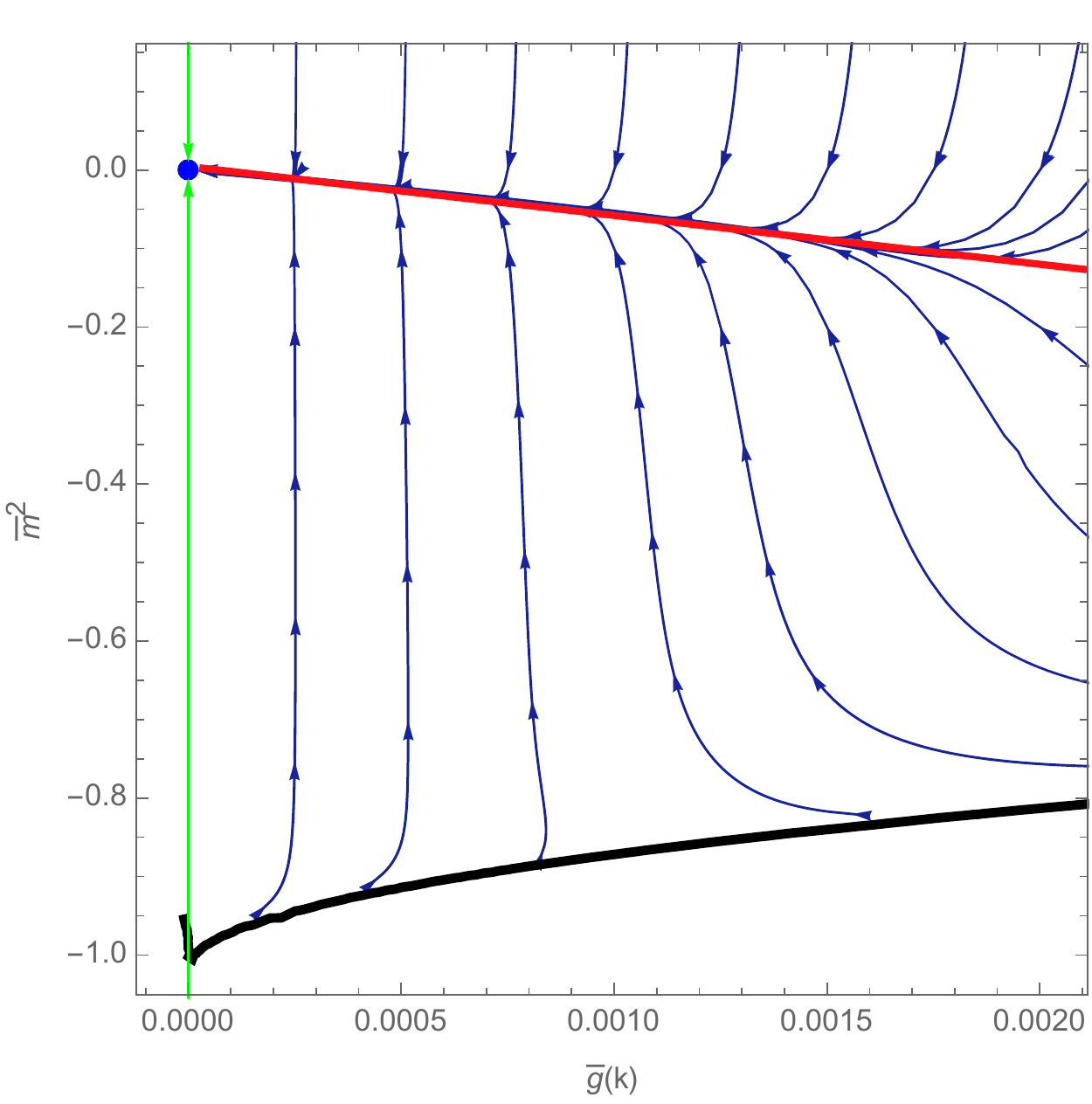} }}}$
\caption{The numerical renormalization group flow arround the Gaussian fixed point, for $\bar{g}(g)<0$ (a) and $\bar{g}(k)>0$ (b). The blue point on both sides corresponds to the Gaussian fixed point, whereas the red point on (a) corresponds to the fixed point $p_1$. The black line correspond the singularity $\mathrm{den}=0$, and the green line corresponds to eigendirections of the Gaussian fixed point. regular trajectories are pictured in blue. }\label{PLOT}
\end{figure}

\noindent
The system \eqref{systprime} completed with the new anomalous dimension \eqref{etaprime} both describe the physical RG flow over $\mathcal{E}_{\mathcal{C}}$. Note that setting $\bar{m}^2\to 0$ and keeping only the first order contributions in $\bar{g}$, we get:
\begin{equation}
\beta_g\approx -\eta \bar{g}\,,\quad\,\beta_m\approx-2\bar{m}^2\,,\quad \eta\approx 4\pi^2\bar{g}\,,
\end{equation}
recovering the well known asymptotic freedom. As expected, the same result may be obtained from the unconstrained system \eqref{syst3}, or from a direct perturbative computation. As a result, the physical space $\mathcal{E}_{\mathcal{C}}$ is connected to the Gaussian fixed point $(\bar{g},\bar{m}^2)=(0,0)$. The flow equations has essentially two source of singularity. The first one for $\bar{m}^2=-1$ due to the symmetric phase restriction, and the second one due to the denominator of $\eta$, $\mathrm{den}(\bar{g},\bar{m}^2):=(1+\bar{m}^2)^5-\Omega^\prime$. From a direct inspection, the Gaussian fixed point is into the region $\mathrm{den}>0$, and the relevant investigated region have to satisfy $\bar{m}^>-1$ and $\mathrm{den}>0$. Numerical investigations show that there are no global fixed points over $\mathcal{E}_{\mathcal{C}}$ for the global fixed point. Indeed, we get three non-Gaussian fixed points: $p_1\approx(1.25,-0.13)$, $p_2\approx (-9,6.6)$ and $p_3\approx(-0.9, 0.0006)$. The two last ones are in the region $\mathrm{den}<0$, and therefore disconnected from the Gaussian fixed point. For $p_1$ however $\mathrm{den}(p_1)>0$. This fixed point has zero anomalous dimensions $\eta(p_1)=0$ and two relevant directions; with critical exponents $(\theta_1,\theta_2)\approx(-4.4, -0.3)$ and eigendirections :
\begin{equation}
\textbf{v}_1\approx(-1,0; -0.1)\,,\qquad \textbf{v}_2\approx (0.9, -0.2)\,.
\end{equation}
Figures \ref{PLOT}a and \ref{PLOT}b describes respectively the behavior constrained RG flow for $\bar{g}\leq 0$ and $\bar{g}\geq 0$ from a numerical integration.\\

\noindent
In contrast with standard analysis based on truncation or unconstrained FRG method like the EVE expansion, there are no global fixed point in the region \ref{PLOT}b. Recalling that all the RG trajectories are oriented from IR to UV, we recognize the Gaussian fixed point as en UV attractor for $\bar{g}(k)>0$, with a very clear large river effect. All the trajectories reach the main stream corresponding to the red line, and finally go to the Gaussian Fixed point. Reversing the arrows, we see that all the trajectories split in two type : the ones going to a region with negative mass and the others, reaching a region with positive mass.

\noindent
This splitting scenario uncontrolled by a fixed point (except the Gaussian one) is reminiscent to a first order phase transition rather than a second order one, as frequently suggested (see \cite{Lahoche:2018hou}-\cite{Eichhorn:2018phj} and  \cite{Sokal:1994un} about first order phase transition). Note the presence of the black singularity line on both sides (a) and (b). In the purely EVE expansion, this singularity have been avoidable, being displaced under the singularity $\bar{m}^2=-1$ from its original position coming from truncations. The resurgence of this singularity is understood as the mark of a significant limitation of our construction, focused on the symmetric phase. Going beyond the symmetric phase and other approximations like \eqref{derivexp}, and investigated the nature of the transition are works in progress. \\

\noindent
On the left hand side (Figure \ref{PLOT}b), for $\bar{g}(k)<0$, the scenario is repeated. The non-Gaussian fixed point $p_1$ behaves like an attracor, with very similar characteristics like the Gaussian fixed point. We have a main stream on both sides of the fixed points, and all the trajectories  reach the stream before to go on $p_1$. The integral curve of the eigendirections for the Gaussian fixed point (in green) separate the flow. Any trajectory on the side $\bar{g}(k)>$ cannot reach the region $\bar{g}(k)<$ and so one. As a result, at least into the investigated region of the phase space, the two regions are disconnected from the RG flow. As a consequence, requiring the coupling to be positive, to ensure integrability of the partition function, it is tempting to view the region \ref{PLOT}b as a formal artifact, and keep only the region \ref{PLOT}b for physical investigation on the considered model.

\section{Conclusion}\label{conclusion}
In this paper we provided a short presentation of an improved version of the standard EVE method, allowing to build an approximation of the exact renormalization group flow sector by sector for a tensorial group field theory, and taking into account systematically the constraint coming from Ward identities along the flow. The resulting effective equations has a single non-trivial fixed point with zero anomalous dimension, positive mass, negative effective coupling and two purely attractive eigendirections. This fixed points, with very similar characteristics like the Gaussian one has been discarded, as the negative coupling region because any trajectories starting from the positive region can reach the negative one. In particular, the melonic constrained flow has no Wilson-Fisher type fixed points, in accordance with the results of our previous works \cite{Lahoche:2018hou}-\cite{Lahoche:2018vun}, and therefore no second order phase transition. Our final landscape is a Gaussian fixed point with a repulsive stream line in the IR.

\section*{Acknowledgements} 
The authors thank Gaba Yae Ulrich for careful reading the end version of the manuscript and for  English spelling check.
\\

\end{document}